\documentclass{elsartL}
\usepackage{graphicx}
\usepackage{amssymb}
\usepackage{amsmath}
\usepackage{mathrsfs}
\begin{document}

\begin{frontmatter}
\title{Proposals for the test of the isospin invariance in the pion-nucleon interaction at low energy}
\author{Evangelos Matsinos},
\author{G\"unther Rasche}

\begin{abstract}
This work elaborates on former remarks of ours regarding two sets of predictions for the observables of the charge-exchange reaction $\pi^- p \to \pi^0 n$ at low energy (pion laboratory kinetic energy $T \leq 100$ MeV). The 
first prediction is obtained via the triangle identity from the results of fits to low-energy $\pi^\pm p$ elastic-scattering data, whereas the second is based on the same analysis of the combined low-energy $\pi^+ p$ and 
charge-exchange databases. Assuming the integrity of the data used in our fits (i.e., the absence of significant systematic effects in the determination of the absolute normalisation of the datasets) and the insignificance 
of residual effects in the electromagnetic corrections, a significant difference between these two sets of predictions may be interpreted as departure from the isospin invariance in the low-energy $\pi N$ interaction. We 
examine the sensitivity of the standard low-energy observables to the effect, and identify the kinematical regions which are promising for experimental survey. Accurate experiments under the suggested conditions will 
differentiate between the predictions, and thus provide an independent test of the isospin invariance in the low-energy $\pi N$ interaction.\\
\noindent {\it PACS 2010:} 13.75.Gx; 25.80.Dj; 25.80.Gn; 11.30.-j
\end{abstract}
\end{frontmatter}

\section{\label{sec:Intro}Introduction}

Were the isospin invariance fulfilled in the hadronic part of the pion-nucleon ($\pi N$) interaction, two (complex) scattering amplitudes (namely the isospin $I=3/2$ amplitude $f_3$ and the $I=1/2$ amplitude $f_1$) would suffice 
in accounting for the three low-energy $\pi N$ reactions, i.e., for the two elastic-scattering (ES) reactions $\pi^\pm p \to \pi^\pm p$ and for the $\pi^- p$ charge-exchange (CX) reaction $\pi^- p \to \pi^0 n$. In that case, 
the $\pi^+ p$ reaction would involve $f_3$, whereas the $\pi^- p$ ES and CX reactions would be described by the linear combinations $(2 f_1 + f_3)/3$ and $\sqrt{2} (f_3-f_1)/3$, respectively. Evidently, the following expression 
(known as `triangle identity') relates the hadronic amplitudes $f_{\pi^+ p}$, $f_{\pi^- p}$, and $f_{\rm CX}$:
\begin{equation} \label{eq:EQ001}
f_{\pi^+ p} - f_{\pi^- p} = \sqrt{2} f_{\rm CX} \, \, \, .
\end{equation}

The isospin invariance in the low-energy $\pi N$ interaction (pion laboratory kinetic energy $T \leq 100$ MeV) was addressed in several works over the past $25$ years \cite{gibbs1995,matsinos1997a,matsinos1997b,matsinos2006,matsinos2013,PSAUpdate}. 
Assuming the integrity of the input data used therein (i.e., the absence of significant systematic effects in the determination of the absolute normalisation of the datasets) and the insignificance of residual effects in the 
electromagnetic (EM) corrections, these studies established isospin breaking, and agreed well among themselves regarding the size of the `anomaly' at low energy, reporting a $5 - 10$ \% effect in the scattering amplitude. 
Contrary to the findings of these studies, calculations conducted within the framework of the heavy-baryon Chiral-Perturbation Theory \cite{hoferichter2010} placed the isospin-breaking effects in the $\pi N$ interaction 
around the $1$ \% level.

A number of approaches for the investigation of the phenomenon by means of analyses of the low-energy $\pi N$ measurements have been put forward.
\begin{itemize}
\item In Ref.~\cite{gibbs1995}, the $\pi^- p$ CX scattering amplitude $f_{\rm CX}$, obtained from the (very few, at that time when the paper appeared) $\pi^- p$ CX measurements, was compared with corresponding predictions 
obtained via Eq.~(\ref{eq:EQ001}) from an analysis of the two ES reactions.
\item References \cite{matsinos1997a,matsinos1997b,matsinos2006,matsinos2013} rested upon a comparison between the $\pi^- p$ CX measurements and corresponding predictions for the low-energy $\pi^- p$ CX observables - i.e., 
for the differential cross section (DCS), for the analysing power (AP), and for the total cross section (TCS) - obtained from the fitted values of the parameters of the hadronic model of this programme (ETH model) and from 
the Hessian matrices of the fits to ES measurements.
\item A third approach, the one this work relates to, was implemented a few years ago \cite{PSAUpdate}, featuring the comparison between two sets of predictions for the observables of the $\pi^- p$ CX reaction~\footnote{As 
the parameterisation of the standard spin-isospin $s$- and $p$-wave phase shifts lies at the basis of our modelling, we need to combine the DBs of at least two low-energy $\pi N$ reactions in order to determine both isospin 
amplitudes ($f_1$ and $f_3$). This restriction does not necessarily apply to other approaches, e.g., to the method of Ref.~\cite{gibbs1995}, which can determine the scattering amplitude $f_{\rm CX}$ directly from the $\pi^- p$ 
CX measurements alone.}: one based on fits to the ES database (DB), denoted henceforth as (DB$_{+/-}$), and the other extracted from fits to the $\pi^+ p$ and the $\pi^- p$ CX measurements, comprising a DB which we denote as 
DB$_{+/0}$. Significant differences between these two sets of predictions may be interpreted as evidence of the violation of the isospin invariance in the low-energy $\pi N$ interaction.
\end{itemize}

The question arises as to which of the three low-energy $\pi^- p$ CX observables is best suited to differentiating between the two sets of predictions of Ref.~\cite{PSAUpdate}. Also relevant in this context is the kinematical 
region, i.e., the ($T$,$\theta$) domain - $\theta$ being the centre-of-momentum (CM) scattering angle - which is best suited to differentiating between the predictions for that observable. These two questions (i.e., best-suited 
observable, best-suited kinematical region) comprise the subject of this work. The hope is that this short note will direct any future experimental activity towards a reduced kinematical region for the most promising 
observable(s).

But why are new $\pi^- p$ CX measurements called for? Why should one not split the available low-energy DB$_0$ into two subsets, namely measurements which would be used in the fits and measurements which would be used in the 
hypothesis testing, and proceed with the proposed test? We argue as follows. To start with, any such splitting of the existing DB$_0$ would be arbitrary; in addition, due to the different sensitivity of the low-energy $\pi^- p$ 
CX measurements to the effect under investigation, one could influence the result of the test by skilfully manipulating the two subsets of the DB$_0$. Any such approach would be tainted as `hypothesis testing under foreknowledge', 
a controversial subject in statistical analyses. One could randomly select the two subsets from the pool of the available experiments, but the chances are that (regardless of the result of the test) any outcome would be prone 
to criticism. Although it is rather unclear at the present time where they could be conducted, we believe that new measurements are indispensable in order to settle down the subject of the isospin invariance in the low-energy 
$\pi N$ interaction. To reliably confirm or refute the validity of an effect observed in a statistical analysis of data, the recommended approach is to `reset the time', acquire new measurements, and investigate whether that 
effect persists.

For the sake of brevity, a prediction or a set of predictions, obtained from the results of the fits to the DB$_{+/-}$, will be named `prediction(s) A' in the following; similarly, a prediction or a set of predictions, obtained 
from the results of the fits to the DB$_{+/0}$, will be named `prediction(s) B'.

\section{\label{sec:Method}Method}

The details about the fitting procedure, including the experimental input, can be found in Ref.~\cite{PSAUpdate}, ZRH19 solution (version v2). Predictions for the $\pi^- p$ CX DCS and AP will be obtained on a ($T$,$\theta$) 
grid, where $T$ will be varied between $10$ and $100$ MeV (with an increment of $5$ MeV) and $\theta$ between $0$ and $180^\circ$ (with an increment of $5^\circ$). At each grid point, one million Monte-Carlo (MC) events will 
be generated for each $m_\sigma$ mass (see Ref.~\cite{PSAUpdate}), using the fitted values of the parameters of the ETH model and the Hessian matrix of each fit. As always, the MC generation will be based on the standard 
routines CORSET and CORGEN of the CERN Program Library, see item V122 in Ref.~\cite{cernlib}: the former evaluates the `square root' of the Hessian matrix, which is required as input to the latter; the output of CORGEN in 
each MC event is a set of correlated normally-distributed random numbers, which lead to the model-parameter vector for that event.

Corresponding (i.e., involving the same ($T$,$\theta$) grid point) predictions $v_1$ and $v_2$ are compared by means of two quantities. The first quantity measures the compatibility of the two values: the absolute normalised 
difference of $v_1$ and $v_2$ is defined as
\begin{equation} \label{eq:EQ002}
d = \frac{\lvert v_1 - v_2 \rvert}{\sqrt{(\delta v_1)^2 + (\delta v_2)^2}} \, \, \, ,
\end{equation}
where $\delta v_1$ and $\delta v_2$ denote the root-mean-square (rms) uncertainties of $v_1$ and $v_2$, respectively. In our recent analyses, we adopted the $2.5 \sigma$ level (in the normal distribution) as the outset of 
statistical significance. Also in this work, two predictions $v_1$ and $v_2$ will be considered significantly different if their absolute normalised difference $d$ exceeds $2.5$.

If the uncertainties of the measurements, on which the hypothesis testing relies, could be made arbitrarily small, then the only quantity needed in this report would be the absolute normalised difference of Eq.~(\ref{eq:EQ002}); 
however, this is hardly the case in scattering measurements. The dominant source of systematic uncertainty in the low-energy $\pi N$ experimentation is associated with the normalisation uncertainty of the datasets: at the 
present time, the average normalisation uncertainty (over the reported results) of the $\pi^- p$ CX DCS experiments at low energy is about $5.8$ \%. To be able to differentiate between the predictions A and B with confidence, 
an experiment must have a normalisation uncertainty which is significantly smaller than the difference between the predictions. Making use of the significance level of $2.5 \sigma$, it follows that an experiment (measuring 
a $\pi^- p$ CX observable) with the current average normalisation uncertainty can differentiate between predictions which are at least $2.5 \times 5.8$ \% apart, i.e., more than about $15$ \% apart. In summary, a significant 
difference between two predictions is not the only issue; the predictions must also be well-separated, so that the experiment have resolving power. (Of course, the statistical uncertainty of the measurements is also relevant. 
Regarding this point, the assumption is that the data acquisition spans a temporal interval extensive enough so that the systematic effects - i.e., the normalisation uncertainty - become dominant.)

Consequently, a second quantity needs to be introduced into the study: the `symmetrised relative difference' (also known as `symmetric absolute percentage error') between the predictions $v_1$ and $v_2$, defined as
\begin{equation} \label{eq:EQ003}
w = 2 \frac{\lvert v_1 - v_2 \rvert}{\lvert v_1 \rvert + \lvert v_2 \rvert} \, \, \, .
\end{equation}
(In this work, $v_1, v_2 \geq 0$ in all cases. The value of $0$ will be assigned to $w$ when $v_1=v_2=0$, as the case is for the AP at $\theta=0$ and $180^\circ$.)

Maps of the quantity $w$ on the ($T$,$\theta$) grid for those of the values $v_1$ and $v_2$ which are significantly different ($d \geq 2.5$) are expected to reveal which of the $\pi^- p$ CX observables are effective in 
differentiating between the predictions A and B, and to identify the kinematical regions which are promising in the hypothesis testing proposed in this work.

Predictions for the $\pi^- p$ CX TCS will also be obtained between $T=10$ and $100$ MeV, with an increment of $5$ MeV. Finally, one additional quantity, as potentially effective in differentiating between the two sets of 
predictions, namely the position (i.e., the $T$ value) of the $\pi^- p$ CX $s$- and $p$-wave interference minimum (destructive interference of the $s$- and the $p$-wave parts of $f_{\rm CX}$), will be considered.

\section{\label{sec:Results}Results}

\begin{figure}
\begin{center}
\includegraphics [width=15.5cm] {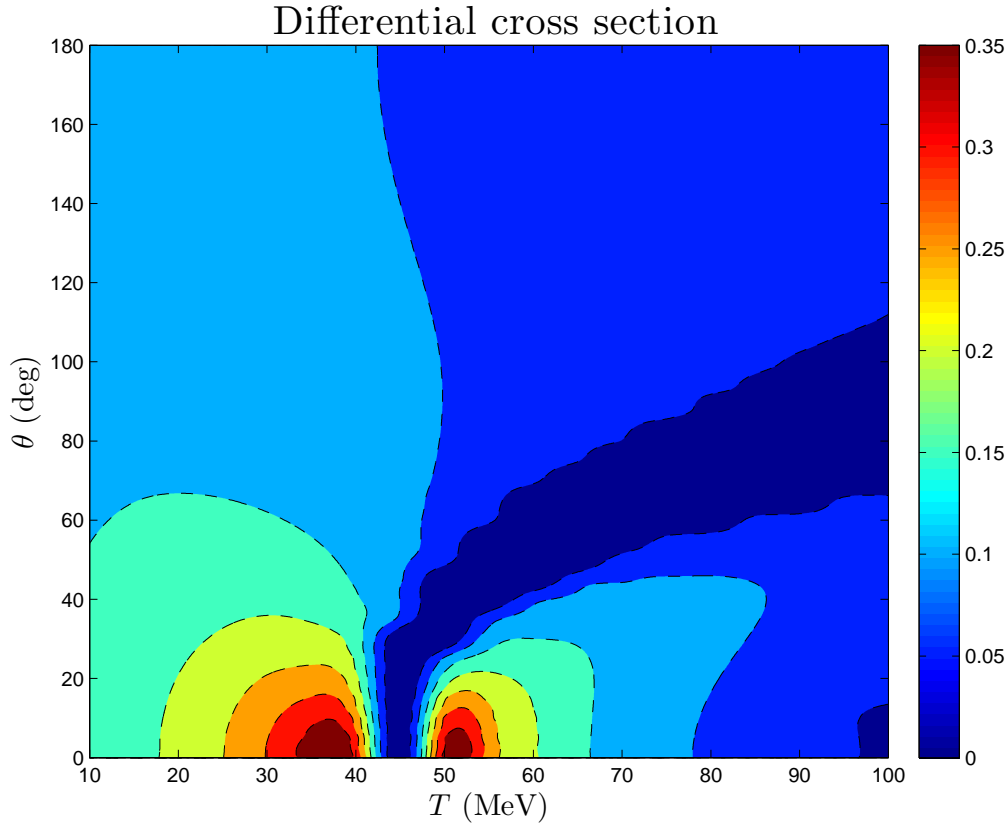}
\caption{\label{fig:CXPredictionsDCS}Map of the quantity $w$ of Eq.~(\ref{eq:EQ003}), representing a measure of separability of the two sets of predictions for the $\pi^- p$ CX DCS for $10 \leq T \leq 100$ MeV and 
$0 \leq \theta \leq 180^\circ$. The promising kinematical regions correspond to maximal $w$ values. The difference between the predictions is statistically significant almost everywhere on the ($T$,$\theta$) grid. For the 
$66$ (out of $703$) ($T$,$\theta$) grid points where $d<2.5$, the quantity $w$ of Eq.~(\ref{eq:EQ003}) was set to $0$.}
\vspace{0.35cm}
\end{center}
\end{figure}

The results for the $\pi^- p$ CX DCS are displayed in Fig.~\ref{fig:CXPredictionsDCS}. The difference between the predictions A and B is significant almost everywhere on the ($T$,$\theta$) grid. The promising kinematical 
region involves the maximisation of $w$: this occurs in forward scattering ($\theta \lesssim 30^\circ$) at energies neighbouring the $s$- and $p$-wave interference minimum. As a matter of fact, this region overlaps with the 
($T$,$\theta$) domain explored in the FITZGERALD86 experiment \cite{fitzgerald1986}. Despite the fact that such an experiment is anything but easy (the DCS at the $s$- and $p$-wave interference minimum is below $10~\mu$b/sr), 
we recommend the repetition of the FITZGERALD86 experiment. On either side of the minimum, the difference between the two sets of predictions is significant and, furthermore, the predictions are well-separated, so that an 
experiment with the current average (for low-energy $\pi^- p$ CX scattering) normalisation uncertainty will suffice in differentiating between the two sets of predictions. One word of advice: detailed measurements at the 
interference minimum are not expected to be particularly helpful; for the sake of demonstration, one datapoint suffices. It makes more sense to focus on accurate measurements on either side of the minimum, where the 
separability of the two predictions is enticing (e.g., one measurement around $36.5$ MeV, another around $51.5$ MeV).

\begin{figure}
\begin{center}
\includegraphics [width=15.5cm] {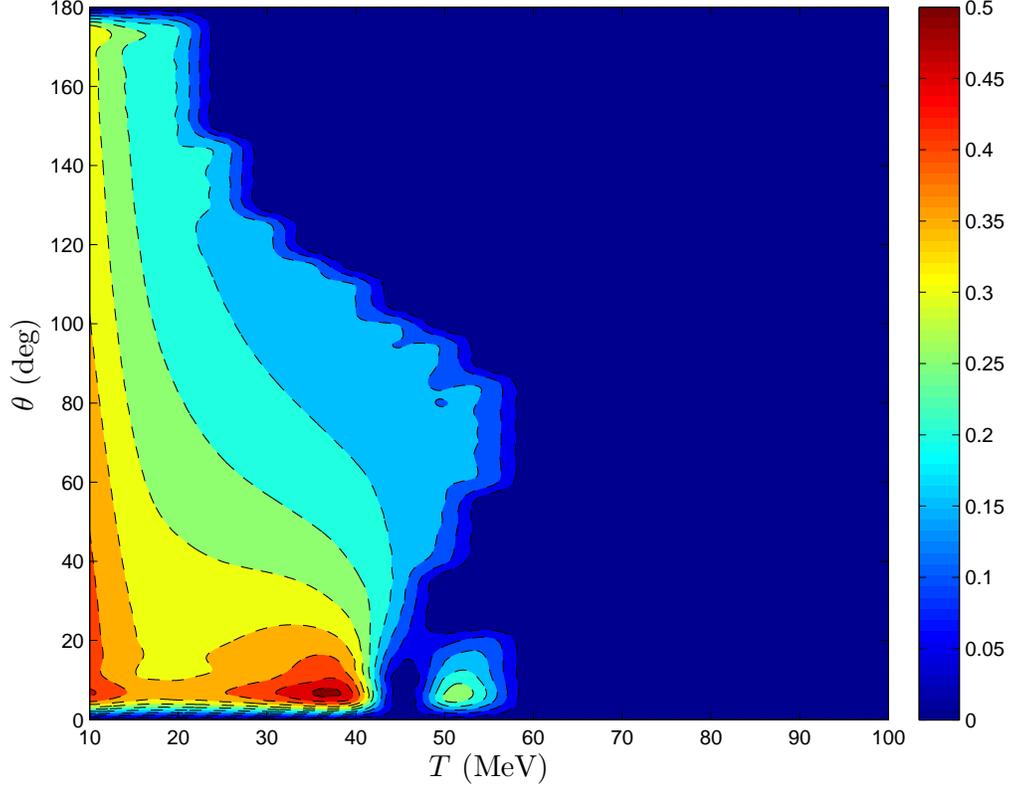}
\caption{\label{fig:CXPredictionsAP}Same as Fig.~\ref{fig:CXPredictionsDCS} for the $\pi^- p$ CX AP. The AP identically vanishes at $\theta=0$ and $180^\circ$.}
\vspace{0.35cm}
\end{center}
\end{figure}

The results for the $\pi^- p$ CX AP are displayed in Fig.~\ref{fig:CXPredictionsAP}. The difference between the predictions A and B is not significant in the larger part of the ($T$,$\theta$) grid. On the other hand, 
measurements within the triangular domain, defined by the ($T$,$\theta$) points: ($10$ MeV,$0^\circ$), ($10$ MeV,$180^\circ$), and ($40$ MeV,$0^\circ$), can differentiate between the two sets of predictions (also notice the 
`island' in forward scattering around $51.5$ MeV). As the AP is essentially a ratio of cross sections, many systematic effects (which find their way into the DCS measurements) drop out. On average, the normalisation 
uncertainties of AP datasets are smaller than those of DCS datasets, and are dominated by the uncertainty in the target polarisation (which is usually around $3$ \%). Unfortunately, very few low-energy measurements of the 
$\pi^- p$ CX AP (a mere ten datapoints) are available at the present time \cite{stasko1993,gaulard1999}; furthermore, these measurements had been acquired close to the maximal $T$ value allowed in our analyses, i.e., in a 
kinematical region which, according to Fig.~\ref{fig:CXPredictionsAP}, is not at all promising in connection with the hypothesis testing proposed in this work.

\begin{figure}
\begin{center}
\includegraphics [width=15.5cm] {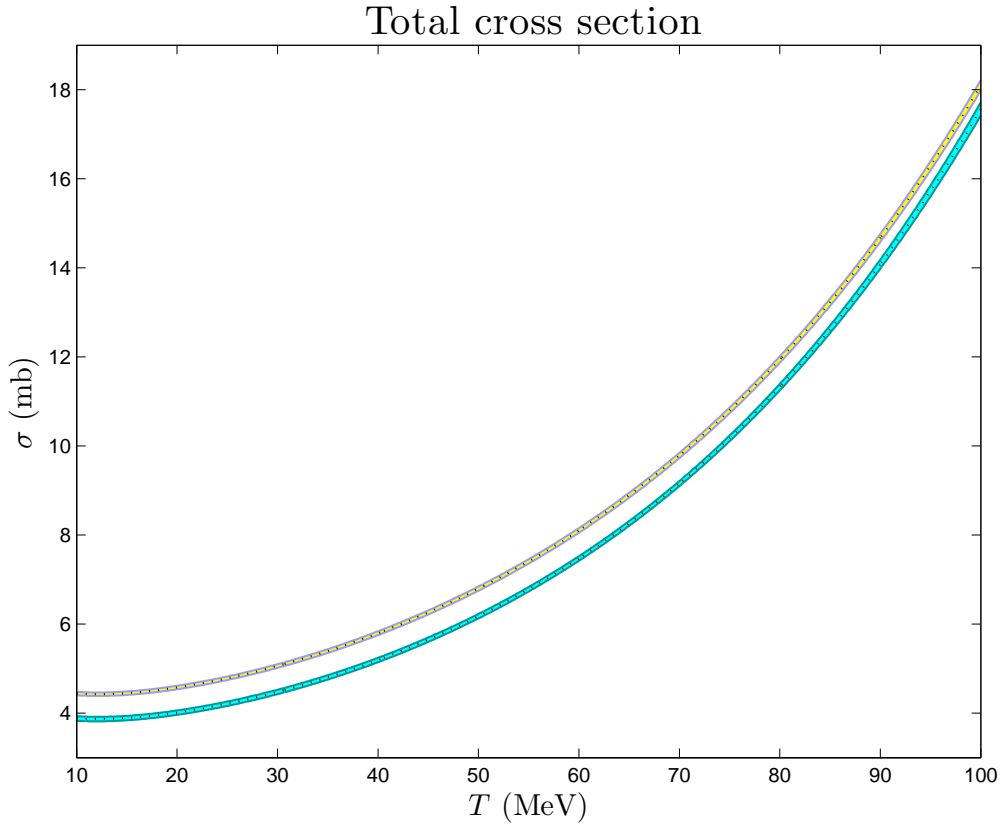}
\caption{\label{fig:CXPredictionsTCS}The two sets of predictions for the $\pi^- p$ CX TCS; lower band: predictions A, upper band: predictions B. The difference between the predictions is statistically significant in the 
entire $T$ domain, yet (owing to the present-day normalisation uncertainty of low-energy $\pi N$ experiments) these sets are not well-separated.}
\vspace{0.35cm}
\end{center}
\end{figure}

The results for the $\pi^- p$ CX TCS are displayed in Fig.~\ref{fig:CXPredictionsTCS}. The difference between the two sets of predictions is statistically significant in the entire $T$ domain, yet the predictions are not 
well-separated. The quantity $w$ of Eq.~(\ref{eq:EQ003}) ranges between about $13.4$ (at $T=10$ MeV) and $3.0$ (at $T=100$ MeV) \%. To be able to differentiate between the predictions, an experiment, measuring the $\pi^- p$ 
CX TCS at $T \approx 10$ MeV, must be accompanied by a normalisation uncertainty of no more than about $5$ \%; this is not trivial. The demands on the normalisation uncertainty at $T=40$ and $95$ MeV, representing the range 
of the $T$ values (below $100$ MeV) of the only relevant recent experiment \cite{breitschopf2006}, are more stringent: $4.4$ and $1.4$ \%, respectively.

\begin{figure}
\begin{center}
\includegraphics [width=15.5cm] {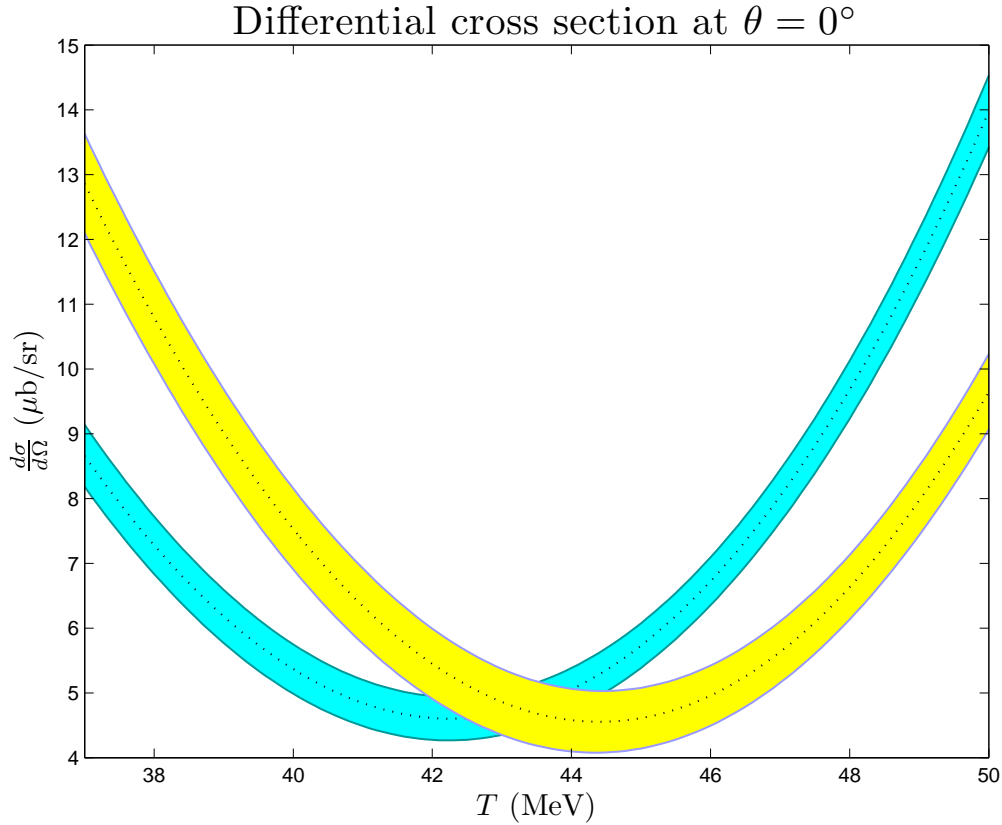}
\caption{\label{fig:CXPredictionsFDCS}The two sets of predictions for the $\pi^- p$ CX DCS at $\theta=0^\circ$ around the $s$- and $p$-wave interference minimum; blue band: predictions A, yellow band: predictions B.}
\vspace{0.35cm}
\end{center}
\end{figure}

The results for the $\pi^- p$ CX DCS at $\theta=0^\circ$ around the $s$- and $p$-wave interference minimum are displayed in Fig.~\ref{fig:CXPredictionsFDCS}. The two predictions disagree regarding the $T$ position of the 
minimum: prediction A places it at $T=42.21(11)$ MeV; prediction B favours $T=44.33(14)$ MeV. (This difference of about $2$ MeV was first addressed in Ref.~\cite{gibbs1995}, see Fig.~3 and the relevant 
text therein.)

\section{\label{sec:Discussion}Discussion}

In a sizeable portion of the kinematical region at low energy, the two sets of predictions obtained for $\pi^- p$ CX observables - namely the one based on the results of fits to the DB$_{+/-}$ (predictions A) and the other 
based on those of the fits to the DB$_{+/0}$ (predictions B) - are significantly different, i.e., different at the significance level assumed in our recent works ($2.5 \sigma$ effect in the normal distribution, corresponding 
to a p-value of about $1.24 \cdot 10^{-2}$). To be able to differentiate between these predictions, and thus provide an independent test of the isospin invariance in the $\pi N$ interaction below $T=100$ MeV, an experiment 
must have sufficient resolving power. At the present time, the average normalisation uncertainty of $\pi^- p$ CX datasets at low energy is about $5.8$ \%. Therefore, to be reliably differentiated by an experiment with this 
normalisation uncertainty, the predictions A and B must be separated by more than about $15$ \%.

Figures \ref{fig:CXPredictionsDCS} and \ref{fig:CXPredictionsAP} show maps of the symmetrised relative difference $w$ of Eq.~(\ref{eq:EQ003}) between the two sets of predictions for the DCS and for the AP, respectively; these 
two plots should be useful in the preparation of experiments aiming at the test of the isospin invariance in the low-energy $\pi N$ interaction. Figure \ref{fig:CXPredictionsTCS} shows the predictions A and B for measurements 
of the TCS. One additional quantity has been explored, namely the $T$ position of the $s$- and $p$-wave interference minimum: the two predictions for the DCS at $\theta=0^\circ$ are displayed in Fig.~\ref{fig:CXPredictionsFDCS}.

Assuming the integrity of the input data used in our fits (i.e., the absence of significant systematic effects in the determination of the absolute normalisation of the datasets) and the insignificance of residual effects in 
the EM corrections, the interpretation of the result of the hypothesis testing, proposed in this work, is straightforward.
\begin{itemize}
\item Compatibility of new experimental results for $\pi^- p$ CX observables with prediction A and incompatibility with prediction B adds a question mark to the issue of the isospin breaking in the low-energy $\pi N$ interaction.
\item The opposite, i.e., incompatibility with prediction A and compatibility with prediction B, provides support for the isospin-breaking scenario, as promulgated in Refs.~\cite{gibbs1995,matsinos1997a,matsinos1997b,matsinos2006,matsinos2013,PSAUpdate}.
\end{itemize}

Predictions for the low-energy $\pi N$ observables (DCS, AP, and TCS) for the three $\pi N$ reactions are simple to obtain, free of charge, and available within a few days of a request. Unlike the predictions obtained from 
dispersion relations, our estimates are accompanied by uncertainties which reflect the statistical and systematic fluctuation of the fitted experimental data.

The figures of this paper have been created with MATLAB$^{\textregistered}$ (The MathWorks, Inc., Natick, Massachusetts, United States).

\end{document}